\documentclass[10pt,twocolumn,letterpaper]{article}

\usepackage{cvpr}
\usepackage{times}
\usepackage{epsfig}
\usepackage{graphicx}
\usepackage{amsmath}
\usepackage{amssymb}
\usepackage[short]{optidef}
\usepackage{subfig}
\usepackage{fancyhdr}
\fancypagestyle{page1}
{
   \fancyhf{}

   \chead{CVPR Challenge on Learned Image Compression (CLIC) Workshop, Salt Lake City, Utah, USA, June 2018}

}
\usepackage[breaklinks=true,bookmarks=false]{hyperref}

\cvprfinalcopy 

\begin{document}

\title{Learned Compression Artifact Removal by Deep Residual Networks \vspace{-14pt}}

\author{Ogun Kirmemis \qquad Gonca Bakar \qquad A. Murat Tekalp\\
 {Department of Electrical and Electronics Engineering, Koc University, 34450 Istanbul, Turkey}\\
 {\tt\small \{okirmemis16,gbakar15,mtekalp\}@ku.edu.tr  \vspace{-10pt}}
}

\maketitle
\thispagestyle{page1}
\begin{abstract} \vspace{-10pt}
We propose a method for learned compression artifact removal by post-processing of BPG compressed images. We trained three networks of different sizes. We encoded input images using BPG with different QP values. We submitted the best combination of test images, encoded with different QP and post-processed by one of three networks, which satisfy the file size and decode time constraints imposed by the Challenge. The selection of the best combination is posed as an integer programming problem. Although the visual improvements in image quality is impressive, the average PSNR improvement for the results is about 0.5 dB.
\end{abstract}
\vspace{-22pt}

\section{Introduction}
\label{sec:intro}
\vspace{-2pt}
The mainstream approach for lossy image compression since 1980's has been transform coding, using discrete cosine transform (DCT) or discrete wavelet transform (DWT) for data decorrelation followed by uniform quantization and entropy coding. The JPEG standard using the DCT has been the most successful and widely deployed lossy image compression technology. JPEG2000, which uses the DWT, is the technology used by the motion picture industry for frame by frame compression of movies.

Recently, the state of the art in lossy image coding has shifted to the better portable graphics (BPG) codec \cite{bpg}, which is also a transform coder derived from intra-frame coding tools in the high-efficiency video coding (HEVC) video coder. The royalty-free WebP codec, which is derived from the intra-frame coding tools of the VP9 video coder, also outperforms JPEG but is slightly inferior to BPG.

With the advent of deep learning, which led to significant achievements in computer vision, there is growing interest in applying end-to-end deep learning to image compression. Many works have already been published on novel encoder/decoder architectures, learned transforms, and learning to better quantize real variables \cite{theis} \cite{rippel} \cite{balle} \cite{dumas} \cite{toderici} \cite{johnston} \cite{covell} \cite{li} \cite{santurkar}.

Our hypothesis in this paper is that end-to-end learned image compression methods have not yet matured to the level to beat the state of the art signal-processing-based transform codecs, e.g., the BPG codec. Hence, we propose a learned post-processing method to improve the visual perceptual quality of BPG compressed images. 


\section{Related Works}
\label{sec:related}
\vspace{-2pt}
Available post-processing methods can be classified as traditional filters and learned artifact reduction methods.

Traditional filters for removal of compression artifacts include deblocking and deringing filters that were proposed as in-loop or post-processing filters to be used with image/video compression standards. An example for in-loop filters is the HEVC deblocking filter \cite{hevc-debl}. Commonly used post-processing filters are those of Foi {\it et al.}~\cite{foi}, which proposed thresholding in shape-adaptive DCT domain for deblocking; 
Zhang {\it et al.}~\cite{zhang}, which proposed similarity priors for image blocks to reduce compression artifacts by estimating the transform coefficients of overlapped blocks from non-local blocks; and
Dar {\it et al.}~\cite{yehuda}, which modeled the compression-decompression procedure as a linear system and then estimate the solution to the inverse problem.

Methods using deep learning for post-processing of compression artifacts include
Dong {\it et al.}~\cite{dong}, which proposes 4 layer convolutional neural network for deblocking and deblurring of compressed images;
Svoboda {\it et al.}~\cite{svoboda}, which proposes an 8 layer residual network and add a loss term defined by the difference between the first partial derivatives of the target and output images to the MSE loss; and
Galteri {\it et al.}~\cite{galteri}, which proposes a solution based on generative adversarial networks (GAN) to reduce compression artifacts.
\vspace{-10pt}


\section{System and Network Architecture}
\label{sec:architectures}
\begin{figure}[h]
\begin{minipage}[b]{1.0\linewidth}
  \centering
\centerline{\epsfig{figure=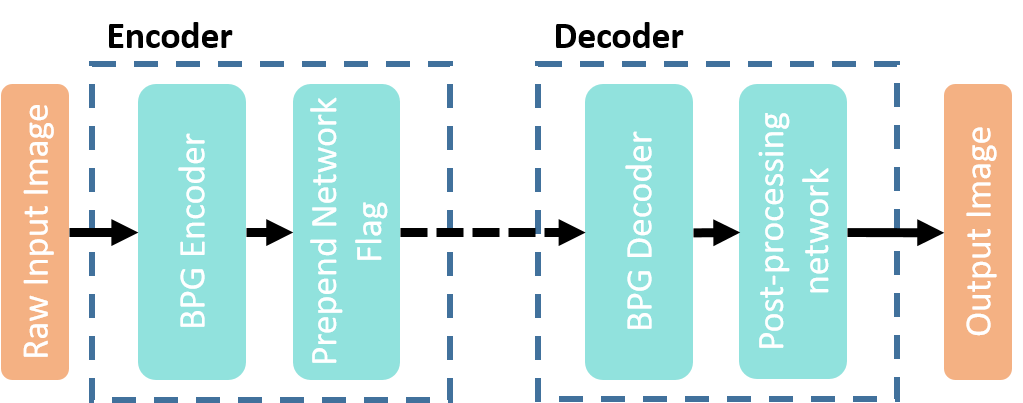,width=7cm}}
\vspace{-3pt}
\end{minipage}
\caption{Block diagram of encoding/decoding system.}
\label{fig:overall}
\vspace{-10pt}
\end{figure}

\vspace{-3pt}
The proposed system is depicted in Figure~\ref{fig:overall}. The encoder unit uses the BPG encoder \cite{bpg}. The decoder unit comprises of a BGP decoder and post-processing network. Since we trained three different networks, the encoder adds a byte in the beginning of the bit-stream to signal the choice of the neural network. Decoder unit reads the first byte and sends the rest of the bitstream to the BPG decoder. Then, the decompressed image is processed by the selected post-processing network yielding the final output image. 

\begin{figure}[h]
\begin{minipage}[b]{1.0\linewidth}
  \centering
\centerline{\epsfig{figure=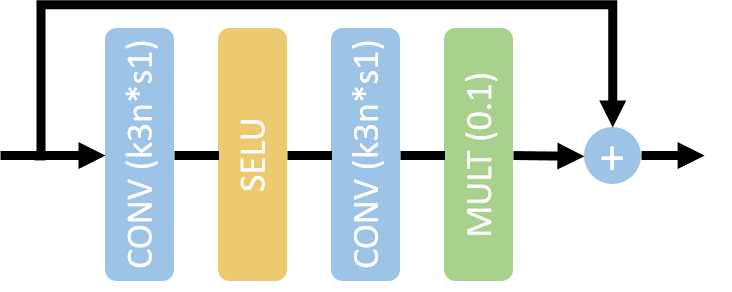,width=7cm}}
\vspace{-5pt}
\end{minipage}
\caption{A residual block with kernel size {\it k}, number of feature maps {\it n} and stride {\it s}. We also employ residual scaling to make training easier as suggested in \cite{resscale}.}
\label{fig:resblock}
\vspace{-6pt}
\end{figure}

The proposed neural network is a modified version of the enhanced deep super-resolution (EDSR) network \cite{edsr}, which is based on SRResNet \cite{srgan} architecture. The main difference between EDSR and ours is that we use SELU activation function \cite{selu} instead of ReLU as shown in Figure~\ref{fig:resblock}, since SELU activation enables faster learning \cite{selu}. We also remove the upsampling blocks of SRResNet. Unlike the~networks in \cite{edsr} and \cite{srgan}, we add a direct shortcut connection from the input RGB image to output RGB image. Since our aim is to restore compressed images, the input image is closer to the output image than the randomly initialized network from the point of optimization. Because of this, we also multiply the contribution of the network with $0.1$. This way the overall function for the network is closer to identity function so that the predictions of the network resemble the input image at the beginning of training.

\begin{figure}[h]
\begin{minipage}[b]{1.0\linewidth}
  \centering
\centerline{\epsfig{figure=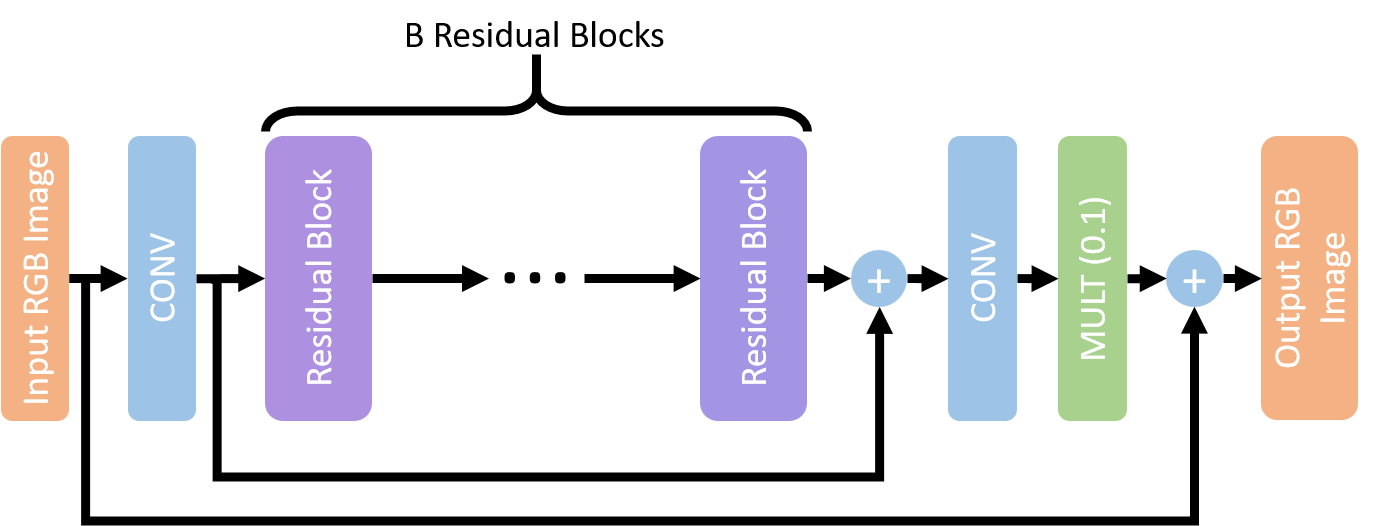,width=7cm}}
\vspace{-3pt}
\end{minipage}
\caption{Architecture of the proposed post-processing network with {\it B} residual blocks. There is a direct shortcut connection from the input image to output image.}
\label{fig:network}
\vspace{-6pt}
\end{figure}

In order to comply with RAM requirements (8 GB) on the evaluation server, our decoder divides the input image to 4 blocks and processes these blocks separately. We employ an Overlap-Save method, to produce the same output as if the whole image is processed at once. In order to apply the Overlap-Save method, the effective kernel size of the neural network has to be calculated. For a network which has $l$ convolutional layers with kernel size of $k$, the effective kernel size $E$ of the overall network is $E=(k-1)l+1$. After we divide the input image to 4 blocks, we pad each block on all sides to size $\frac{E+1}{2}$. Then, we pass these blocks through the network. When merging the output blocks, we discard overlapping pixels and construct the output image.


\section{Training Method}
\label{sec:training}
We train 3 models with different depth $B$, referring to the number of blocks, and width $n$, referring to the number of feature maps in Figures~\ref{fig:network} and \ref{fig:resblock}. These networks are called MVGL A (B=32, n=96) with 5.40M parameters, MVGL B (B=8, n=96) with 1.42M parameters, and MVGL C (B=32, n=48) with 1.35M parameters. MVGL A is trained with batch size of 24, while both MVGL B and MVGL C are trained with batch size of 64.

We train all networks with the given training set consisting of 1633 images. We encode the training images using the BPG encoder with QP=40 at the highest compression level (9). QP=40 is the minimum QP value that we can choose to fit the validation set into the given constraint of 0.15 bits per pixel (bpp). 

We calculate the mean of the RGB channels of all images in the training set (single mean per channel for the training set), and subtract them from both target images and their compressed/decompressed versions before feeding them into the network. We train networks on MSE loss using the Adam optimizer\cite{Adam} with the default parameters ($\beta_{1}=0.9$, $\beta_{2}=0.999$). The learning rate is initialized to $0.001$ and is halved at every $500^{th}$ epoch.  Networks are trained on $96\times 96$ random crops without any data augmentation. A random patch of size $96\times 96$ is cropped randomly from every training image to create training batch for an epoch. We stop training networks upon convergence, that is, when there is no improvement for 50 epochs.
\vspace{3pt}


\section{Evaluation}
\label{sec:experiments}

We present PSNR and MS-SSIM results for different QP and networks on the given training, validation, and test sets. 

\subsection{Results on Training and Validation Sets}
The average PSNR and MS-SSIM results on the training and validation sets encoded with QP=40 are shown in Table \ref{tab:qp40res}. MVGL A is the best performing network with PSNR gain of $\approx$ 0.7dB on both training and validation sets, since it has the largest number of parameters (weights). MVGL~B and MVGL C networks give comparable results with $\approx$~0.3-0.4~dB PSNR improvements, since the number of parameters in these networks are close to each other. 
\vspace{2pt}

\begin{table}[h]
\begin{center}
\caption{Results on the training set (0.169 bpp) and validation set (0.149 bpp) where QP=40 for BPG compression. \vspace{-23pt}} 
\vspace{6pt}
\label{tab:qp40res}
\begin{tabular}{|c|c|c|c|c|c|}
\hline
\multicolumn{1}{|c|}{} & \multicolumn{2}{|c|} {Training Set} & \multicolumn{2}{|c|}{Validation Set}\\
\cline{2-5}
\hline
Method & PSNR & MS-SSIM & PSNR & MS-SSIM\\
\hline
BPG    & 30.529 & 0.948 & 30.842 & 0.948\\
\hline
MVGL A & 31.221 & 0.955 & 31.533 & 0.955\\
\hline
MVGL B & 30.899 & 0.951 & 31.223 & 0.952\\
\hline
MVGL C & 30.952 & 0.952 & 31.277 & 0.950\\
\hline
\end{tabular}
\end{center}
\vspace{-12pt}
\end{table}

\subsection{Encoding the Test Set with File Size Constraint}
Suppose there are $N$ images in the test set and we need to choose the best QP value of out of $M$ different values for each image to maximize the average PSNR of BPG encoding subject to a file size constraint.  We formulate this~problem as an integer linear programming problem, given by
\vspace{-16pt}
\begin{mini!}[2]
{x_i}{ \sum_{i=1}^{N} f_i^T x_i \label{eq:obj}}{\label{eq:optprob}}{}
\addConstraint{ \sum_{i=1}^{n} b_{i}^{T} x_i }{\leq FileSizeLimit \label{eq:size}}
\addConstraint{ \hspace{-16pt} \mathbf{1}_{1\times M} x_{i}  }{=1, \; \forall i=1,2,\cdots N \label{eq:select}}
\addConstraint{\hspace{-16pt} x_{i_{j}} \in \left \{ 0,1 \right \}, }{\; \forall i=1,2,\cdots N,\; \forall j=1,2,\cdots M 
\label{eq:int}}
\vspace{-8pt}
\end{mini!}
where $x_i$ is $M\times 1$ one-hot vector such that the entry which equals 1 indicates the QP selected for the $i^{th}$ image, $f_i$ is $M\times 1$ vector whose components are the sum of squared error between the raw and encoded images for different QPs, and $b_i$ is $M\times1$ vector whose components denote the file size when $i^{th}$ image is encoded with all possible QP. Eqn.~\ref{eq:size} enforces that the sum of sizes of all images are below the given file size constraint $FileSizeLimit$. Constraints \ref{eq:select} and \ref{eq:int} require that only one QP is selected for each image. 

We solved this problem for $N=286$ and $M=5$ corresponding to QP values 38 to 42. Solution of this problem reveals that we should encode 1 image with QP=38, 109 images with QP=40, 120 images with QP=41, and 56 images with QP=42 so that the average bitrate is 0.15 bpp. 

\subsection{Results on the Test Set}
\vspace{-2pt}
We encoded $N=286$ images in the test set with the QP values determined above to meet the file size constraint. We now need to determine which of the three networks to use for post-processing of each image. Our results in Table 2 indicate that the best results can be obtained by the network MVGL A; however, the total processing time was too long. MVGL B and MVGL C are considerably faster but yield lower PSNR improvements. 

{\bf Submitted Results}: Because the training of MVGL C was not complete by the Challenge submission deadline, we decided to process 71 images by MVGL A and the remaining images by MVGL B so that the total processing time is less than $\approx$ 45 hours on Intel i7-3630QM 2.40GHz CPU. This method is called MVGL in Table 2 and is our submission to the Challenge. 71 images to be processed by MVGL A are selected such that they yield the biggest PSNR improvement when processed by MVGL A instead of MVGL B. Had we considered combination of MVGL A and MVGL C for our submission, the average PSNR would be 30.180.

\begin{table}[h]
\begin{center}
\caption{Average PSNR and MS-SSIM and time (mins) for the Test Set. The average bit-rate is 0.15 bpp. PSNR gain is the difference between PSNR of post-processed and BPG. \vspace{-12pt}} 
\vspace{6pt}
\label{tab:results}
\begin{tabular}{|c|c|c|c|c|}
\hline
Method & PSNR & PSNR Gain & MS-SSIM & Time\\
\hline
BPG    & 29.692 & -     & 0.944 & \\
\hline
MVGL A & 30.267 & 0.575 & 0.950 & 6210\\
\hline
MVGL B & 30.011 & 0.319 & 0.947 & 1582\\
\hline
MVGL C & 30.052 & 0.360& 0.947 & 634\\
\hline
\textbf{MVGL} & 30.135 & 0.443 & 0.948 & 2725\\
\hline
\end{tabular}
\end{center}
\vspace{-20pt}
\end{table}

\begin{figure*}[tb]
\vspace{-10pt}
\centering
\subfloat[Image \# (rectangle shows crop location)]{\epsfig{figure=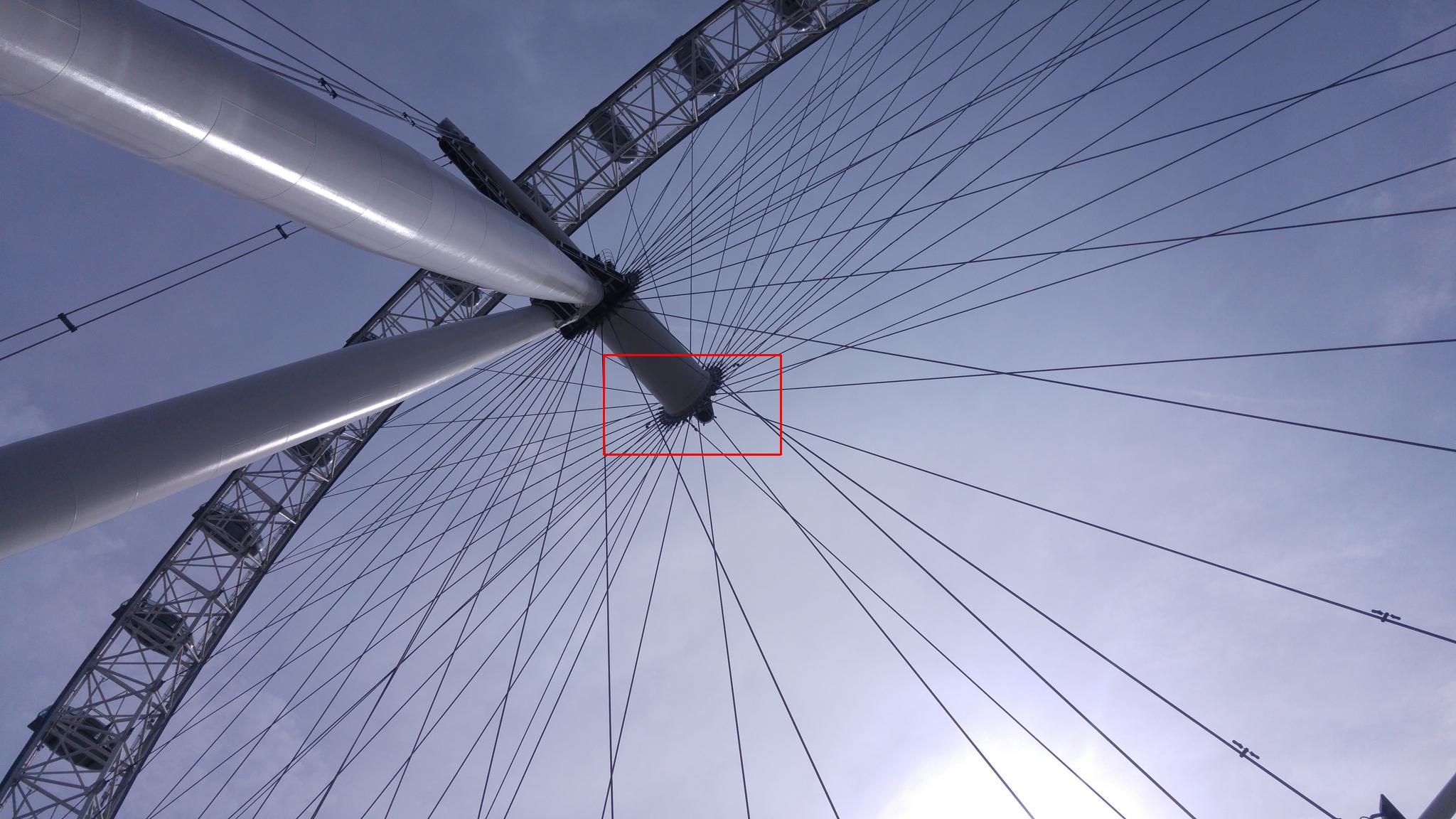,width=5cm}} \quad
\subfloat[BPG (crop) , PSNR=32.417 dB]{\epsfig{figure=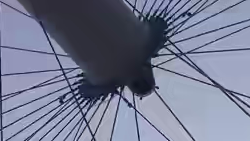,width=5cm}} \quad
\subfloat[MVGL A processed, PSNR=34.801 dB]{\epsfig{figure=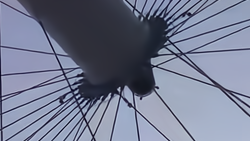,width=5cm}} \vspace{-10pt} \\
\subfloat[Image \# (rectangle shows crop location)]{\epsfig{figure=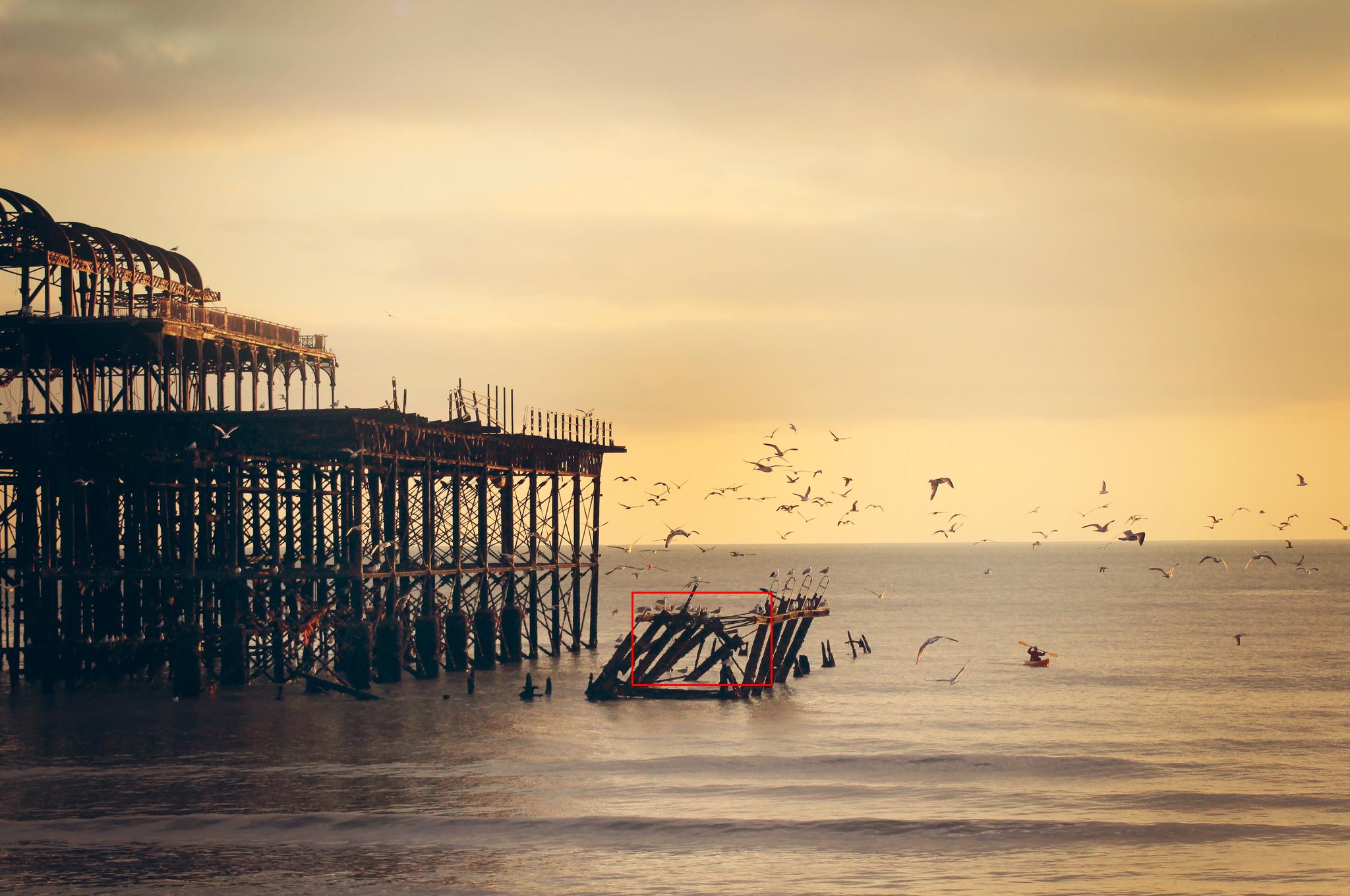,width=5cm}} \quad
\subfloat[BPG (crop), PSNR=31.798 dB]{\epsfig{figure=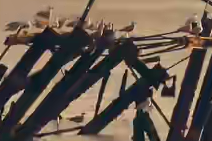,width=5cm}} \quad
\subfloat[MVGL A processed, PSNR=33.445 dB]{\epsfig{figure=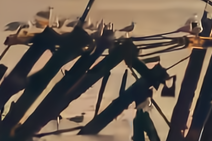,width=5cm}} \vspace{-6pt}
\caption{Visual results for two images from the Test Set. }
\vspace{-8pt}
\label{fig:visual}
\end{figure*}

{\bf Complete Results}:
Table~\ref{tab:qp} presents average PSNR and MS-SSIM values for all combinations of encoding all images in the test set with QP values 39-43 and post-processing them by all three networks. In each row of the table, we encode all images in the test set with the same QP. Table \ref{tab:qp} shows that all three networks provide solid PSNR gains across different QP values which means that the networks generalize for different QP values well even though they are only trained with images encoded using QP=40. All images show impressive visual quality improvement. Two example visual results are shown in Figure \ref{fig:visual}.
\vspace{-2pt}

\begin{table*}[tb]
\begin{center}
\caption{Results for the test set encoded with different QP values. \vspace{-6pt}} 
\label{tab:qp}
\begin{tabular}{|c|c|c|c|c|c|c|c|c|c|}
\hline
\multicolumn{1}{|c|}{} & \multicolumn{2}{|c|} {BPG} & \multicolumn{2}{|c|}{MVGL A} & \multicolumn{2}{|c|}{MVGL B} &  \multicolumn{2}{|c|}{MVGL C} & \multicolumn{1}{|c|}{}\\
\cline{2-5}
\hline
QP & PSNR & MS-SSIM & PSNR & MS-SSIM & PSNR & MS-SSIM  & PSNR & MS-SSIM & Bitrate (bpp)\\
\hline
39 & 30.833 & 0.954 & 31.486 & 0.960 & 31.189 & 0.957 & 31.238 & 0.957 & 0.206\\
\hline
40 & 30.333 & 0.949 & 30.956 & 0.955 & 30.674 & 0.952 & 30.720 & 0.953 & 0.179\\
\hline
41 & 29.735 & 0.943 & 30.334 & 0.950 & 30.064 & 0.946 & 30.108 & 0.947 & 0.152\\
\hline
42 & 29.249 & 0.938 & 29.802 & 0.944 & 29.554 & 0.941 & 29.594 & 0.941 & 0.132\\
\hline
43 & 28.687 & 0.930 & 29.200 & 0.937 & 28.970 & 0.934 & 29.008 & 0.934 & 0.111\\
\hline
\end{tabular}
\end{center}
\vspace{-10pt}
\end{table*}

\vspace{-3pt}
\section{Conclusions}
\label{sec:conclusion}
\vspace{-3pt}
The success of the proposed deep learning methods for post-processing of compressed/decompressed images depends on availability of sufficient processing power for both training and testing. Our results (comparing Network B and Network C) show that the average PSNR over the test set (for the same rate) improves by the depth of the network. However, the computational load of test phase with even moderately deep networks can be demanding. As a result, we were not able to submit our best results for the challenge, but only those results that conform with the computational constraints imposed by the Challenge administrators.
\vspace{-3pt}
{\small
\bibliographystyle{ieee}
\bibliography{egpaper_final}
}

\end{document}